%% file: main.tex
\documentclass[conference]{IEEEtran}
\IEEEoverridecommandlockouts

\usepackage{amsmath,amssymb}
\usepackage{booktabs}
\usepackage{graphicx}
\usepackage{tabularx}
\usepackage{array}
\usepackage{cite}
\usepackage{url}
\usepackage{dblfloatfix}
\usepackage{placeins}
\graphicspath{{figures/main/}{figures/optional/}}

\newcommand{\CVaR}{\operatorname{CVaR}}
\newcommand{\RCVaR}{R_{\mathrm{CVaR}}}

\title{Regime-Dependent Value of CVaR in Preventive Maintenance Scheduling under RUL Uncertainty
\thanks{Funded by the European Union. Views and opinion expressed are however those of the author(s) only and do not necessarily reflect those of the European Union or Europe’s Rail Joint Undertaking. Neither the European Union nor the granting authority can be held responsible for them. The project FP3-IAM4Rail is supported by the Europe's Rail Joint Undertaking and its members.}
}

\author{%
\IEEEauthorblockN{Jerzy Baranowski and Waldemar Bauer}
\IEEEauthorblockA{Department of Automatic Control and Robotics\\
AGH University of Science and Technology\\
Krakow, Poland\\
\{jb,bauer\}@agh.edu.pl}
}

\begin{document}
\maketitle

\begin{abstract}
We study preventive-maintenance scheduling for a small fleet when several assets compete for a limited number of maintenance slots and their remaining useful lives (RULs) are uncertain. The planning horizon is divided into discrete periods; each asset may be maintained at most once, and at most $K$ assets may be maintained in any one period. Future usage and RUL errors are represented by scenarios. We compare a risk-neutral policy that minimizes expected cost with a risk-aware policy that minimizes expected cost plus conditional value-at-risk (CVaR), where $\CVaR_{0.90}$ is the mean cost among the worst 10\% of scenarios. Both policies are solved exactly by enumerating all capacity-feasible schedules. A $3\times5$ experiment combines capacities $K=1,2,3$ with five increasing levels of RUL uncertainty, with ten paired replications for each combination. The risk-aware policy provides its largest benefit when RUL uncertainty is low-to-moderate and more than one maintenance action can be executed per period. In the best observed case, it reduces CVaR by 13.4\% with an expected-cost increase of only 0.2 relative cost units. At high uncertainty, the benefit becomes negligible and the two policies can select identical schedules. The results show that risk-aware scheduling creates operational value only when prognostic information is sufficiently informative and the maintenance system has enough flexibility to act on it.
\end{abstract}


\input{manuscript/sections/01_introduction}
\input{manuscript/sections/02_problem_formulation}
\input{manuscript/sections/03_experimental_design}
\input{manuscript/sections/04_results}
\input{manuscript/sections/05_discussion}
\input{manuscript/sections/06_conclusion}

\FloatBarrier
\bibliographystyle{IEEEtran}
\bibliography{referencesICTMOD}

\end{document}

%% file: manuscript/sections/01_introduction.tex
\section{Introduction}

Predictive maintenance converts condition-monitoring and prognostic information into decisions about when individual assets should be serviced. In a fleet, these decisions cannot always be made independently: several assets may require intervention at similar times, while the available crew, workshop, or production window can accommodate only a limited number of actions. The planner must therefore decide not only whether an asset should be maintained, but also how the available maintenance slots should be allocated over the planning horizon. This decision is uncertain because the remaining useful life (RUL) supplied by a prognostic model is an estimate rather than a known failure time.

An expected-cost objective provides a natural basis for such planning, but it can underweight rare and expensive outcomes. For example, two schedules may have almost the same average cost while one produces much larger losses in a small fraction of adverse scenarios. Conditional value-at-risk (CVaR) provides a way to include these upper-tail costs directly in the scheduling objective. However, adding a risk term does not necessarily change the selected schedule. Its value depends on whether the RUL estimates distinguish the urgent assets reliably and whether the capacity constraints leave alternative feasible allocations.

Maintenance optimization has progressed from fixed time- and age-based rules toward condition-based and prognostics-supported decision making~\cite{Ahmad2012,deJonge2020,Bousdekis2019,Li2020}. RUL predictions have been incorporated into maintenance scheduling~\cite{Lei2018,Shi2025}, while stochastic multi-component formulations account for uncertain degradation, interacting assets, and shared resources~\cite{OldeKeizer2017,Zhu2021,IslamVatn2023,Einabadi2023}. Risk-averse maintenance policies further recognize that decision makers may seek protection from unfavorable cost realizations rather than minimize expected cost alone~\cite{PedersenVatn2022,Rockafellar2000}. What remains less clear is when an explicit tail-risk term produces a materially different and operationally better schedule.

Our preceding study introduced a finite-horizon scenario-based framework that combines calendar limits, uncertain future usage, and uncertain RUL in one maintenance-schedule evaluation problem~\cite{baranowski2026optimizationpredictivemaintenanceschedules}. Its proof-of-concept example showed that integrated policies can strongly outperform single-trigger rules, but the expected-cost and CVaR-based schedules were almost identical. That formulation did not include a shared per-period maintenance-capacity constraint. The present study extends the framework by making assets compete for a limited number of maintenance slots and by varying both the capacity and the uncertainty of the RUL estimates.

The research question is: \emph{How do RUL uncertainty and maintenance capacity determine whether a CVaR-aware objective changes maintenance schedules and reduces tail operational cost?} The contributions are as follows:
\begin{itemize}
    \item We formulate a capacity-constrained scheduling problem in which each preventive-maintenance action occupies one of a limited number of slots available in a planning period.
    \item We compare a risk-neutral expected-cost policy with a mean--CVaR policy over the same feasible schedules and solve both problems by exhaustive enumeration.
    \item We conduct a controlled $3\times5$ experiment combining three capacity levels with five increasing levels of RUL uncertainty.
    \item We identify when the risk-aware policy reduces upper-tail cost at negligible average-cost premium and when uncertainty or lack of capacity prevents it from improving the schedule.
\end{itemize}

The rest of the paper is organized as follows. Section~II defines the scheduling decisions, the maintenance-capacity constraint, the scenario cost, and the two optimization criteria. Section~III describes the capacity--uncertainty experiment, the exhaustive solution procedure, and the paired out-of-sample evaluation. Section~IV presents the numerical results. Section~V discusses their operational interpretation and limitations, and Section~VI concludes the paper.

%% file: manuscript/sections/02_problem_formulation.tex
\section{Scheduling Model and Risk Criteria}

\subsection{Maintenance decisions and capacity}

Consider a fleet of assets $\mathcal I=\{1,\ldots,N\}$ and a planning horizon divided into periods $\mathcal T=\{1,\ldots,T\}$. A period may represent, for example, a day, week, or maintenance window. The binary variable
\begin{equation}
    x_{i,t}=
    \begin{cases}
        1, & \text{if asset $i$ is maintained in period $t$,}\\
        0, & \text{otherwise}
    \end{cases}
    \label{eq:xit}
\end{equation}
defines the schedule. Each asset may receive at most one preventive-maintenance action during the horizon,
\begin{equation}
    \sum_{t\in\mathcal T} x_{i,t} \leq 1, \qquad i\in\mathcal I.
    \label{eq:at-most-once}
\end{equation}
If asset $i$ is maintained, its selected period is denoted by $\tau_i$; the value $\tau_i=0$ denotes that no preventive action is scheduled within the horizon.

The maintenance capacity $K$ is the maximum number of actions that can be completed in one period. Every action uses one homogeneous maintenance slot, hence
\begin{equation}
    \sum_{i\in\mathcal I} x_{i,t} \leq K, \qquad t\in\mathcal T.
    \label{eq:capacity}
\end{equation}
Thus, $K=1$ permits only one asset to be maintained in a given period, whereas $K=3$ permits up to three assets to be maintained in that period. Capacity is renewed in the next period; it is not a limit on the total number of actions over the entire horizon. For example, the schedule $(\tau_1,\tau_2,\tau_3,\tau_4)=(2,2,5,0)$ is feasible for $K\geq 2$ but infeasible for $K=1$, because two assets are assigned to period 2.

The set of feasible schedules is
\begin{equation}
    \mathcal X_K=
    \left\{x\in\{0,1\}^{N\times T}: \eqref{eq:at-most-once} \text{ and } \eqref{eq:capacity} \text{ hold}\right\}.
    \label{eq:feasible-set}
\end{equation}
The capacity constraint couples the asset decisions: assigning one asset to a crowded period can force another asset to be maintained earlier, later, or not within the horizon.

\subsection{Asset information and uncertain futures}

Each asset is described by three maintenance-relevant quantities inherited from the scenario-based framework in~\cite{baranowski2026optimizationpredictivemaintenanceschedules}: calendar age, accumulated usage, and an estimated remaining useful life (RUL). Calendar and usage limits indicate when intervention becomes due under conventional rules. Future usage is uncertain, and the RUL estimate contains prognostic error.

A scenario $\xi_s$ is one possible joint future over the horizon. It specifies the future usage and the effective remaining life of every asset. The five RUL-uncertainty settings used later are generated by increasing the dispersion of the RUL estimation error while leaving its central value unchanged. Consequently, a higher uncertainty level does not mean that the assets are expected to fail sooner; it means that their estimated urgency is less reliable. Under high uncertainty, two assets that appear to have different RULs may have similar or reversed actual failure times.

For schedule $x$ and scenario $\xi_s$, the simulator returns the total cost
\begin{equation}
    C(x,\xi_s)=\sum_{i\in\mathcal I}\left(C_i^{\mathrm{PM}}(x,\xi_s)+C_i^{\mathrm{F}}(x,\xi_s)\right).
    \label{eq:scenario-cost}
\end{equation}
The planned-maintenance term $C_i^{\mathrm{PM}}$ accounts for performing the preventive action at the selected time, including the modelled consequence of intervening earlier than necessary. The failure term $C_i^{\mathrm{F}}$ is incurred when the scenario-specific condition of the asset reaches the failure threshold before its scheduled intervention. If no preventive action is scheduled, the asset remains exposed until the end of the horizon. Costs are reported in relative cost units. The scheduling problem therefore balances the cost of early intervention against the potentially much larger cost of intervening too late.

\subsection{Expected cost, VaR, and CVaR}

For a fixed schedule, the scenario cost $C$ is a random loss. Its value-at-risk at confidence level $\alpha\in(0,1)$ is
\begin{equation}
    \operatorname{VaR}_{\alpha}(C)=\inf\{z\in\mathbb{R}: \Pr(C\leq z)\geq \alpha\}.
    \label{eq:var-definition}
\end{equation}
The corresponding conditional value-at-risk is defined through the Rockafellar--Uryasev representation~\cite{Rockafellar2000}
\begin{equation}
    \CVaR_{\alpha}(C)=\min_{\eta\in\mathbb{R}}\left\{\eta+\frac{1}{1-\alpha}\,\mathbb{E}[(C-\eta)_+]\right\},
    \label{eq:cvar-definition}
\end{equation}
where $(u)_+=\max(u,0)$. For a continuous cost distribution, $\CVaR_{\alpha}$ is the mean cost in the worst $1-\alpha$ fraction of outcomes. In this study, $\alpha=0.90$, so CVaR summarizes the average cost among the worst 10\% of scenarios.

For $S$ equiprobable scenarios $\{\xi_s\}_{s=1}^{S}$, the expected cost and CVaR are estimated by
\begin{equation}
    \widehat{\mu}(x)=\frac{1}{S}\sum_{s=1}^{S} C(x,\xi_s),
    \label{eq:sample-mean}
\end{equation}
and
\begin{equation}
    \widehat{\CVaR}_{\alpha}(x)=\min_{\eta\in\mathbb{R}}\left\{\eta+\frac{1}{(1-\alpha)S}\sum_{s=1}^{S}(C(x,\xi_s)-\eta)_+\right\}.
    \label{eq:sample-cvar}
\end{equation}

\subsection{Compared scheduling policies and performance measures}

The risk-neutral policy, denoted P1, selects the feasible schedule with the smallest estimated expected cost,
\begin{equation}
    x^{\star}_{\mathrm{P1}}\in\arg\min_{x\in\mathcal X_K}\widehat{\mu}(x).
    \label{eq:p1}
\end{equation}
The risk-aware policy, denoted P2, augments expected cost with a CVaR penalty,
\begin{equation}
    x^{\star}_{\mathrm{P2}}\in\arg\min_{x\in\mathcal X_K}\left[\widehat{\mu}(x)+\lambda\widehat{\CVaR}_{\alpha}(x)\right],
    \label{eq:p2}
\end{equation}
where $\lambda\geq 0$ determines the weight assigned to upper-tail cost. P1 and P2 use the same assets, scenarios, constraints, and solution method. Their only difference is the objective function.

For replication $r$, the relative CVaR reduction achieved by P2 is
\begin{equation}
    \RCVaR^{(r)}=
    \frac{\CVaR_{\alpha}^{\mathrm{P1},(r)}-\CVaR_{\alpha}^{\mathrm{P2},(r)}}{\CVaR_{\alpha}^{\mathrm{P1},(r)}},
    \label{eq:rcvar}
\end{equation}
so a positive value means that P2 has lower tail cost. The expected-cost premium is
\begin{equation}
    \Delta\mu^{(r)}=\mu^{\mathrm{P2},(r)}-\mu^{\mathrm{P1},(r)},
    \label{eq:dmu}
\end{equation}
where a positive value is the average-cost price paid for risk reduction. Finally,
\begin{equation}
    D^{(r)}=\frac{1}{N}\sum_{i=1}^{N}\mathbb{I}\!\left(\tau_i^{\mathrm{P1},(r)}\neq \tau_i^{\mathrm{P2},(r)}\right)
    \label{eq:D}
\end{equation}
measures the fraction of assets assigned to different maintenance periods by the two policies.

%% file: manuscript/sections/03_experimental_design.tex
\section{Computational Experiment and Solution Procedure}

\subsection{Capacity--uncertainty experiment}

We use a synthetic fleet of $N=4$ assets and a planning horizon of $T=6$ periods. Two factors are varied systematically. The first is the per-period capacity $K\in\{1,2,3\}$, corresponding to one, two, or three maintenance slots in every period. The second is the uncertainty of the RUL estimates, represented by five ordered levels: very low, low-mid, medium, medium-high, and high. All remaining model parameters are kept fixed so that the observed changes can be interpreted in terms of these two factors. The CVaR confidence level is $\alpha=0.90$, and the CVaR weight in P2 is $\lambda=0.5$.

The Cartesian product of the three capacities and five uncertainty levels gives 15 factor combinations. We call each combination a \emph{configuration}. Each configuration is evaluated in ten independent replications. A replication contains a newly generated asset instance, a scenario sample used to select the schedules, and a separate scenario sample used to evaluate them. P1 and P2 receive exactly the same asset data and random scenarios within a replication, making their comparison paired.

\input{manuscript/tables/table1_study_design}

\subsection{Scenario generation and parameterization}

The uncertainty model follows the scenario-based formulation introduced in~\cite{baranowski2026optimizationpredictivemaintenanceschedules}. For each configuration, the optimizer uses 300 sampled future scenarios, and the selected schedules are then evaluated on an independent set of 2000 out-of-sample scenarios. A scenario specifies future usage and the effective remaining life of every asset over the horizon. The usage process is sampled independently across assets, while the five uncertainty levels are created by increasing the dispersion of the RUL estimation error while keeping its central value fixed. Thus, moving from very low to high uncertainty does not make the fleet intrinsically more degraded; it makes the estimated urgency ordering progressively less reliable.

The scenario cost uses two components. The first is a planned-maintenance cost that captures the effect of performing a preventive intervention at the selected period, including the cost of acting earlier than necessary. The second is a failure-related cost incurred when the scenario-specific asset condition reaches the failure threshold before the scheduled intervention. Failure costs are intentionally much larger than the marginal penalty for early intervention, so the optimization must balance premature maintenance against the possibility of a costly late action.

\subsection{Exhaustive solution of the scheduling problems}

The small problem size allows both optimization problems to be solved exactly. Because each of the four assets may be assigned to one of six periods or left without preventive maintenance, there are at most $(T+1)^N = 7^4 = 2401$ raw assignments before applying the capacity constraint. The solution procedure for each replication is:
\begin{enumerate}
    \item Generate the asset data and the scenarios used for schedule optimization.
    \item Enumerate all vectors $(\tau_1,\ldots,\tau_N)$ with $\tau_i\in\{0,1,\ldots,T\}$.
    \item Remove every assignment that allocates more than $K$ assets to the same period.
    \item For each remaining schedule, calculate its cost in every optimization scenario and then compute $\widehat{\mu}(x)$ and $\widehat{\CVaR}_{\alpha}(x)$.
    \item Select the schedule minimizing~\eqref{eq:p1} for P1 and the schedule minimizing~\eqref{eq:p2} for P2.
\end{enumerate}
No heuristic search, relaxation, or local optimization is used. The selected schedules are therefore global optima of their respective finite-scenario problems over the complete capacity-feasible set $\mathcal{X}_K$.

\subsection{Independent paired evaluation}

The scenarios used to choose a schedule are not reused to report its performance. After P1 and P2 have selected their schedules, both schedules are evaluated on the same independent set of future scenarios. Using identical evaluation scenarios for both policies is a common-random-number design: any observed difference is attributable to the schedules rather than to one policy receiving an easier random sample.

For each replication, we calculate expected cost, CVaR, the relative CVaR reduction in~\eqref{eq:rcvar}, the expected-cost premium in~\eqref{eq:dmu}, and the schedule-difference measure in~\eqref{eq:D}. The values shown in the figures and tables are means over the ten replications unless stated otherwise.

To summarize the strength of the evidence in the capacity--uncertainty map, a configuration is called \emph{favorable} when three conditions hold simultaneously: the mean relative CVaR reduction is at least 2.5\%, P2 has lower CVaR in at least eight of the ten replications, and its mean expected-cost premium does not exceed 1.5 cost units. A configuration is called \emph{adverse} when P2 has higher mean CVaR than P1. All other configurations are called \emph{inconclusive}. These labels summarize this finite experiment; they are not claims that every real system with the same qualitative factor levels will behave identically.

%% file: manuscript/tables/table1_study_design.tex
\begin{table}[t]
\caption{Capacity--uncertainty experiment}
\label{tab:study-design}
\centering
\small
\begin{tabular}{@{}ll@{}}
\toprule
Item & Setting \\
\midrule
Assets and horizon & $N=4$, $T=6$ periods \\
Decision per asset & one period or no PM \\
Resource use & one slot per PM action \\
Capacity $K$ & 1, 2, or 3 slots/period \\
RUL uncertainty & five increasing levels \\
Risk parameters & $\alpha=0.90$, $\lambda=0.5$ \\
Factor combinations & $3\times5=15$ \\
Replications & 10 paired runs/combination \\
Optimization & exhaustive feasible-schedule search \\
Evaluation & independent common scenarios \\
\bottomrule
\end{tabular}
\end{table}

%% file: manuscript/sections/04_results.tex
\section{Results}

\begin{figure*}[!t]
    \centering
    \includegraphics[width=0.95\textwidth]{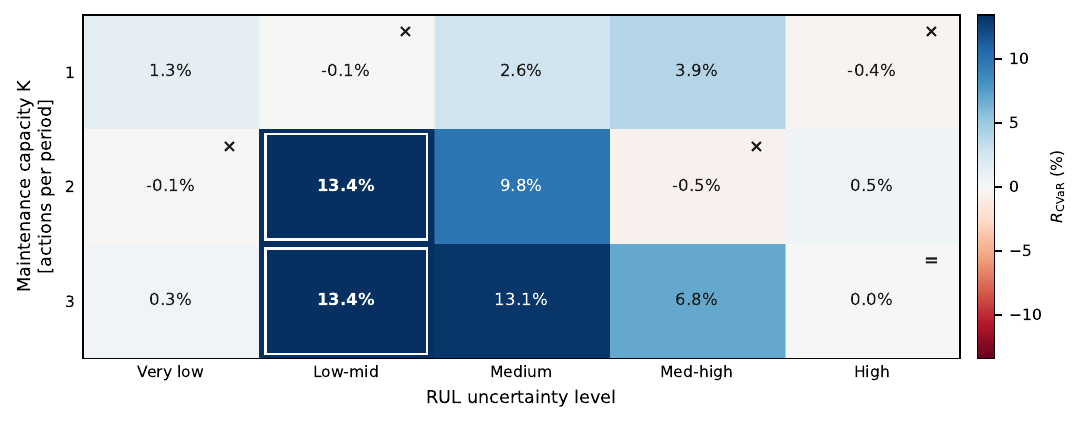}
    \caption{Mean relative CVaR reduction of the risk-aware policy P2 with respect to the expected-cost policy P1 across the $3\times5$ capacity--uncertainty experiment. Rows give the maximum number $K$ of maintenance actions permitted per period, and columns give the RUL-uncertainty level. Cell values report the mean percentage reduction across ten paired replications; positive values favor P2. Outlined cells satisfy all favorable-evidence criteria, $\times$ marks configurations with higher mean CVaR under P2, and $=$ identifies the configuration in which P1 and P2 select identical schedules.}
    \label{fig:stage-b-heatmap}
\end{figure*}

\subsection{Effect of capacity and RUL uncertainty}

Figure~\ref{fig:stage-b-heatmap} reports the mean relative CVaR reduction of the risk-aware policy P2 with respect to the risk-neutral policy P1 for all 15 combinations of capacity and RUL uncertainty. Rows correspond to the maximum number $K$ of maintenance actions allowed in one period, and columns correspond to increasing uncertainty in the RUL estimates. Positive percentages indicate lower CVaR under P2.

The largest reductions occur at low-mid RUL uncertainty with capacities $K=2$ and $K=3$. In both combinations, P2 reduces mean CVaR by 13.4\% relative to P1. At medium uncertainty, positive mean reductions remain visible for $K=2$ and $K=3$, but the replication-level evidence is less consistent and these combinations do not satisfy all favorable-evidence criteria.

The benefit generally weakens as RUL uncertainty increases. At the highest uncertainty level, the mean change ranges from a small deterioration of $-0.4\%$ to an improvement of only $0.5\%$. For $K=3$, P1 and P2 select identical schedules and therefore have identical expected cost and CVaR. The pattern indicates that the CVaR term cannot improve the allocation when the RUL estimates no longer distinguish asset urgency reliably.

Capacity also matters because it controls how many alternative allocations are feasible. With $K=1$, at most one asset can be placed in each period, so the scheduler has little freedom to move several vulnerable assets earlier. Capacities $K=2$ and $K=3$ provide more room for P2 to change the schedule, but this additional flexibility is useful only while the RUL information remains sufficiently informative.

\subsection{Representative factor combinations}

Table~\ref{tab:representative-results} gives three combinations with the same capacity, $K=3$, and progressively increasing RUL uncertainty. Holding capacity fixed makes the effect of prognostic uncertainty easier to interpret. In the low-mid-uncertainty case, P2 reduces CVaR by 13.4\% while increasing expected cost by only 0.2 relative cost units. Its mean schedule-difference measure is $D=0.575$: averaged over the ten replications, 57.5\% of assets are assigned to a different maintenance period than under P1. The tail-risk reduction is therefore associated with substantial rescheduling rather than with two objectives assigning different values to the same decision.

At medium-high uncertainty, the mean CVaR reduction remains positive at 6.8\%, but it is not sufficiently consistent across the ten replications to meet the favorable-evidence rule. At high uncertainty, both policies select the same schedules and all reported differences are zero.

\input{manuscript/tables/table2_representative_results}

\begin{figure*}[!b]
    \centering
    \includegraphics[width=0.95\textwidth]{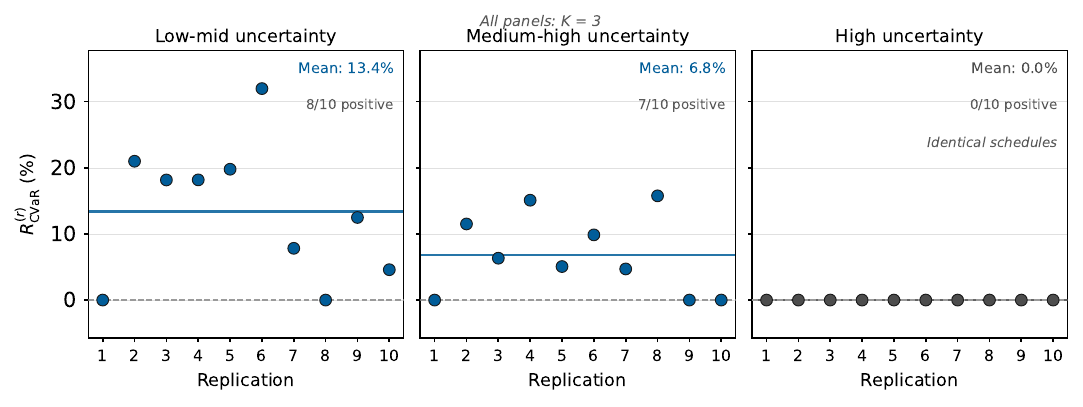}
    \caption{Replication-level relative CVaR reduction of P2 with respect to P1 for three RUL-uncertainty levels at fixed maintenance capacity $K=3$. Each point represents one paired replication, the solid horizontal line gives the mean, and the dashed line marks zero improvement. Panel annotations report the mean reduction and the number of replications with lower CVaR under P2. At high uncertainty, all values are zero because the two policies select identical schedules.}
    \label{fig:representative-profiles}
\end{figure*}

Figure~\ref{fig:representative-profiles} shows the individual replication results for these three combinations. The low-mid-uncertainty case has a consistently positive tendency, the medium-high case contains both stronger and weaker outcomes, and the high-uncertainty case collapses to zero because the schedules are identical.

%% file: manuscript/tables/table2_representative_results.tex
\begin{table}[!t]
\caption{Representative results for P2 relative to P1 at capacity $K=3$}
\label{tab:representative-results}
\centering
\small
\begin{tabular}{lccc}
\hline
\begin{tabular}[c]{@{}l@{}}RUL \\ uncertainty\end{tabular} & \begin{tabular}[c]{@{}c@{}}Mean \\ $\RCVaR$ (\%)\end{tabular} & \begin{tabular}[c]{@{}c@{}}Mean \\ $\Delta E[C]$\end{tabular} & \begin{tabular}[c]{@{}c@{}}Changed \\ assets\end{tabular} \\ \hline
Low-mid                                                    & 13.4                                                          & 0.2                                                           & 0.575                                                     \\
Medium-high                                                & 6.8                                                           & 0.5                                                           & --                                                        \\
High                                                       & 0.0                                                           & 0.0                                                           & 0.000                                                     \\ \hline
\end{tabular}
\end{table}

%% file: manuscript/sections/05_discussion.tex
\section{Discussion}

The results can be interpreted through two requirements for risk-aware scheduling: information and actionability. RUL uncertainty determines the information available to rank the assets. When uncertainty is low or moderate, the scheduler can identify assets for which postponement creates costly upper-tail outcomes. Maintenance capacity determines actionability. When more than one slot is available in a period, the scheduler can move several vulnerable assets earlier without displacing all other interventions.

These requirements are complementary. Additional capacity does not help P2 when the RUL estimates are too uncertain to identify which assets should receive the additional slots. Conversely, accurate RUL estimates have limited decision value when $K=1$ leaves little freedom to change the allocation. The largest reductions therefore occur where reasonably informative prognostics coincide with a non-tight capacity constraint.

The non-monotone pattern across the uncertainty levels is also informative. The strongest effect appears at the low-mid level rather than at the very lowest uncertainty level. A plausible interpretation is that when uncertainty is extremely small, the expected-cost policy P1 already ranks the most urgent assets well, leaving little room for the CVaR term to modify the schedule. At low-mid uncertainty, adverse tail scenarios become more consequential while the urgency ordering is still sufficiently informative, so P2 can improve the allocation. At high uncertainty, this ordering becomes too unreliable for the risk term to help. This explanation is an interpretation of the observed pattern rather than a formal proof of mechanism.

The expected-cost premiums in the representative cases are small because P1 and P2 use the same scenario model and differ only in their treatment of the upper tail. P2 does not abandon average performance; it accepts a small increase in expected cost when moving maintenance earlier substantially reduces costly adverse scenarios. The schedule-difference results show that this trade-off is implemented through different maintenance dates rather than through a purely numerical change in the objective.

Several limitations bound the interpretation. The experiment uses a synthetic fleet of four assets, a six-period horizon, homogeneous unit resource requirements, and at most one preventive action per asset. Exact enumeration is practical at this scale but does not provide a solution method for large fleets. The study also uses ten replications, one CVaR confidence level, 300 optimization scenarios, 2000 evaluation scenarios, and ordered RUL-uncertainty settings. Correlated operating conditions, rolling-horizon re-optimization, repeated maintenance, heterogeneous action durations, and alternative risk levels remain outside the present scope. The results should therefore be read as a controlled demonstration of the capacity--uncertainty mechanism rather than as a calibrated industrial performance claim.

%% file: manuscript/sections/06_conclusion.tex
\section{Conclusion}

We compared expected-cost and mean--CVaR maintenance scheduling for a small fleet subject to a per-period maintenance-capacity constraint. Both policies were solved exactly over the same set of feasible schedules and evaluated on paired out-of-sample scenarios. The risk-aware policy achieved its largest tail-cost reductions when RUL uncertainty was low-to-moderate and at least two maintenance actions could be performed in one period. At high uncertainty, its advantage became negligible and the two policies could select identical schedules. These results show that CVaR-based scheduling is conditionally useful: it requires both prognostic information that distinguishes asset urgency and sufficient maintenance flexibility to act on that information.

%% file: referencesICTMOD.bib
@misc{baranowski2026optimizationpredictivemaintenanceschedules,
  title         = {Optimization of Predictive Maintenance Schedules under Uncertainty: A Scenario-Based Theoretical Framework},
  author        = {Jerzy Baranowski and Waldemar Bauer},
  year          = {2026},
  eprint        = {2605.30222},
  archivePrefix = {arXiv},
  primaryClass  = {eess.SY},
  url           = {https://arxiv.org/abs/2605.30222}
}

@article{Ahmad2012,
  author  = {Rosmaini Ahmad and Shahrul Kamaruddin},
  title   = {An Overview of Time-Based and Condition-Based Maintenance in Industrial Application},
  journal = {Computers \& Industrial Engineering},
  volume  = {63},
  number  = {1},
  pages   = {135--149},
  year    = {2012},
  doi     = {10.1016/j.cie.2012.02.002}
}

@article{deJonge2020,
  author  = {Bram de Jonge and Philip A. Scarf},
  title   = {A Review on Maintenance Optimization},
  journal = {European Journal of Operational Research},
  volume  = {285},
  number  = {3},
  pages   = {805--824},
  year    = {2020},
  doi     = {10.1016/j.ejor.2019.09.047}
}

@article{OldeKeizer2017,
  author  = {Minou C. A. Olde Keizer and Simme Douwe P. Flapper and Ruud H. Teunter},
  title   = {Condition-Based Maintenance Policies for Systems with Multiple Dependent Components: A Review},
  journal = {European Journal of Operational Research},
  volume  = {261},
  number  = {2},
  pages   = {405--420},
  year    = {2017},
  doi     = {10.1016/j.ejor.2017.02.044}
}

@article{Bousdekis2019,
  author  = {Alexandros Bousdekis and Katerina Lepenioti and Dimitris Apostolou and Gregoris Mentzas},
  title   = {Decision Making in Predictive Maintenance: Literature Review and Research Agenda for Industry 4.0},
  journal = {IFAC-PapersOnLine},
  volume  = {52},
  number  = {13},
  pages   = {607--612},
  year    = {2019},
  doi     = {10.1016/j.ifacol.2019.11.226}
}

@article{Li2020,
  author  = {Yanrong Li and Shizhe Peng and Yanting Li and Wei Jiang},
  title   = {A Review of Condition-Based Maintenance: Its Prognostic and Operational Aspects},
  journal = {Frontiers of Engineering Management},
  volume  = {7},
  number  = {3},
  pages   = {323--334},
  year    = {2020},
  doi     = {10.1007/s42524-020-0121-5}
}

@article{Lei2018,
  author  = {Xin Lei and Peter A. Sandborn},
  title   = {Maintenance Scheduling Based on Remaining Useful Life Predictions for Wind Farms Managed Using Power Purchase Agreements},
  journal = {Renewable Energy},
  volume  = {116},
  pages   = {188--198},
  year    = {2018},
  doi     = {10.1016/j.renene.2017.03.053}
}

@article{Zhu2021,
  author  = {Zhicheng Zhu and Yisha Xiang and Bo Zeng},
  title   = {Multicomponent Maintenance Optimization: A Stochastic Programming Approach},
  journal = {INFORMS Journal on Computing},
  volume  = {33},
  number  = {3},
  pages   = {898--914},
  year    = {2021},
  doi     = {10.1287/ijoc.2020.0997}
}

@article{IslamVatn2023,
  author  = {Abu MD Ariful Islam and J{\o}rn Vatn},
  title   = {Condition-Based Multi-Component Maintenance Decision Support under Degradation Uncertainties},
  journal = {International Journal of System Assurance Engineering and Management},
  volume  = {14},
  number  = {4},
  pages   = {961--979},
  year    = {2023},
  doi     = {10.1007/s13198-023-01900-9}
}

@article{PedersenVatn2022,
  author  = {Tom Ivar Pedersen and J{\o}rn Vatn},
  title   = {Optimizing a Condition-Based Maintenance Policy by Taking the Preferences of a Risk-Averse Decision Maker into Account},
  journal = {Reliability Engineering \& System Safety},
  volume  = {228},
  pages   = {108775},
  year    = {2022},
  doi     = {10.1016/j.ress.2022.108775}
}

@article{Einabadi2023,
  author  = {Behnam Einabadi and Mehdi Mahmoodjanloo and Armand Baboli and Eva Rother},
  title   = {Dynamic Predictive and Preventive Maintenance Planning with Failure Risk and Opportunistic Grouping Considerations: A Case Study in the Automotive Industry},
  journal = {Journal of Manufacturing Systems},
  volume  = {69},
  pages   = {292--310},
  year    = {2023},
  doi     = {10.1016/j.jmsy.2023.06.012}
}

@article{Shi2025,
  author  = {Guannan Shi and Xiaohong Zhang and Jianchao Zeng and Haitao Liao and Jie Gan and Jinhe Wang and Zhijian Wang},
  title   = {A Predictive Maintenance Framework Based on Real-Time Credibility Evaluation of Remaining Useful Life Prediction Results},
  journal = {Reliability Engineering \& System Safety},
  volume  = {264},
  pages   = {111342},
  year    = {2025},
  doi     = {10.1016/j.ress.2025.111342}
}

@article{Rockafellar2000,
  author  = {R. Tyrrell Rockafellar and Stanislav Uryasev},
  title   = {Optimization of Conditional Value-at-Risk},
  journal = {Journal of Risk},
  volume  = {2},
  number  = {3},
  pages   = {21--41},
  year    = {2000},
  doi     = {10.21314/JOR.2000.038}
}
